\documentclass[natbib,letterpaper,iop]{emulateapj}
\usepackage{amstext}
\usepackage{apjfonts}
\usepackage{amsmath}
\usepackage{amssymb}
\usepackage[colorlinks=true,linkcolor=blue,citecolor=blue]{hyperref}
\begin{document}

\newcommand{\beq}{\begin{equation}}
\newcommand{\eeq}{\end{equation}}
\newcommand{\beqar}{\begin{eqnarray}}
\newcommand{\eeqar}{\end{eqnarray}}
\newcommand{\Ra}{R_{\rm a}}
\newcommand{\Ma}{M_{\rm a}}
\newcommand{\vinf}{v_\infty}
\newcommand{\rhoinf}{\rho_\infty}
\newcommand{\mach}{\mathcal M_\infty}
\newcommand{\mhl}{\dot M_{\rm HL}}
\newcommand{\lhl}{\dot L_{\rm HL}}
\newcommand{\Hrho}{H_\rho}
\newcommand{\Rs}{R_{\rm s}}
\newcommand{\erho}{\epsilon_\rho}
\newcommand{\cs}{c_{\rm s,\infty}}
\newcommand{\ehl}{\dot E_{\rm HL}}

\title{On the Accretion-Fed Growth of Neutron Stars During Common Envelope } 

\author{Morgan MacLeod and Enrico Ramirez-Ruiz}
\affil{Department of Astronomy and
  Astrophysics, University of California, Santa Cruz, CA
  95064, USA}
   
\begin{abstract} 
This paper models the orbital inspiral of a neutron star (NS) through the envelope of its giant-branch companion during a common envelope (CE) episode. These CE episodes are necessary  to produce close pairs of NSs that can inspiral and merge due to gravitational wave losses in less than a Hubble time. 
Because cooling by neutrinos can be very efficient, NSs have been predicted to accumulate significant mass during CE events, perhaps enough to lead them to collapse to black holes.  
We revisit this conclusion with the additional consideration of CE structure, particularly density gradients across the embedded NS's accretion radius. 
This work is informed by  our recent numerical simulations that find that the presence of a density gradient strongly limits accretion by imposing a net angular momentum to the flow around the NS. 
Our calculations suggest that NSs should survive CE encounters. They accrete only modest amounts of envelope material, $\lesssim 0.1M_\odot$, which is broadly consistent with mass determinations of double NS binaries. 
With less mass gain, NSs must spiral deeper to eject their CE, leading to a potential increase in mergers. 
The survival of NSs in CE events has implications for the formation mechanism of observed double NS binaries, as well as for predicted rates of NS binary gravitational wave inspirals and their electromagnetic counterparts. 
\end{abstract}

\maketitle

\clearpage
\section{Introduction}\label{sec:intro}

The existence of a population of compact neutron star (NS) binaries \citep{Hulse:1975dw} serves as a unique probe of general relativity \citep{Stairs:2004go} and of binary stellar evolution \citep{Bethe:1998jv,Kalogera:2007kh,Postnov:2014fb}. Mergers of NS binaries are promising sources for the detection of gravitational radiation \citep{1991ApJ...380L..17P,Belczynski:2002gi}, and are the progenitors of short gamma ray bursts \citep{Narayan:1992hx,Behroozi:2014bp}. 
Yet, to inspiral and merge under the influence of gravitational radiation in less than a Hubble time, a compact binary must be separated by less than the radii of its main sequence progenitors \citep[e.g.,][]{Peters:1964bc}. To reach their current small separations, these binaries must have passed through one or more common envelope (CE) phases \citep{Paczynski:1976uo}.

In a standard evolutionary scenario to produce NS binaries, the companion to a NS evolves and engulfs the NS inside its growing envelope  \citep{Taam:1978dk,1995ApJ...445..367T,2006csxs.book..623T}. 
Within the shared envelope, the NS focusses envelope gas toward itself. 
Flow convergence leads to dissipation of orbital energy in shocks and to accretion \citep{Hoyle:1939fl,Iben:1993ka,Ivanova:2013co}. 
Neutrinos serve as an effective cooling channel for this convergent flow, allowing  material to be incorporated into the NS at a {\it hypercritical} accretion rate well above the classical Eddington limit \citep{Houck:1991kc,Fryer:1996kr,Popham:1999io,Brown:2000hm,Narayan:2001cm,Lee:2005gp,Lee:2006kp}.
The relative rates of drag and accretion implied by \citet{Hoyle:1939fl} accretion (HLA) theory suggest that an inspiralling NS is likely to grow to collapse to a black hole (BH) before the CE is ejected  \citep{Chevalier:1993by,Armitage:2000eg,BETHE:2007bs}, leaving behind a tightened remnant binary \citep{Webbink:1984jd}.

Despite this apparently clear prediction, reconciling the observed distribution of NS masses  \citep[e.g.,][]{Schwab:2010du,Ozel:2012cv,Kiziltan:2013ky} with theories of hypercritical accretion in CE has posed a long-standing problem. 
In particular, double NSs exhibit a narrow range of inferred  masses  close to the suspected NS birth mass, centered at $1.33M_\odot$  with dispersion of $0.05 M_\odot$ \citep{Ozel:2012cv}.  
Alternative evolutionary scenarios have been proposed where the NS can avoid CE, and accretion, entirely. For example, if the binary is sufficiently equal in mass, it could undergo a simultaneous, or {\it double core}, CE \citep{Brown:1995jj}. 
The issue is that for each binary that passed through a preferred channel one would expect numerous massive NS and BH-NS binaries assembled through the more standard channels  \citep{1998ApJ...502L...9F,Belczynski:2002gi,Kalogera:2007kh,Belczynski:2010ca,2013ApJ...764..181F}. This picture remains at odds with the apparent paucity of BHs just above the maximum NS mass  \citep{Ozel:2010hd,Ozel:2012cv}.

 In this Letter we re-evaluate claims that BHs should necessarily form via
accretion-induced collapse during CE events involving a NS and its massive companion.
We draw on results of our recent simulations
of accretion flows within a stellar envelope to demonstrate that it is critical to consider not just the binding energy, but also the structural properties of the whole envelope  \citep{MacLeod:2014tn}. To this end, we follow the inspiral and accretion of a NS during its dynamical inspiral and show that all NSs should be expected to survive CE evolution, accreting only a small fraction of their own mass.

\section{Characteristic Conditions in NS Accretion}\label{sec:conditions}

When a NS becomes embedded within a CE, it exerts a gravitational influence on its surroundings and can accrete envelope material. In this section, we explore some characteristic scales for that accretion flow, focusing on how they depend on the supply of material and the microphysics of the gas. 

\subsection{Hoyle-Lyttleton Accretion within a CE}\label{sec:CE_HLA}
The flow around the NS can be described in the context of the NS's gravitational interaction with the surrounding medium in HLA theory \citep{Hoyle:1939fl}. The NS's velocity relative to the envelope gas, $\vinf$, is typically mildly supersonic, $\mach = \vinf / \cs \gtrsim 1$, where $\mach$ is the flow Mach number and $\cs$ is the local sound speed. 
Material with an impact parameter less than an accretion radius,
\beq
\Ra = \frac{2 G M_{\rm NS}}{\vinf^2},
\eeq
is focused toward the NS. The resulting accretion rate is
\beq\label{mhl}
\mhl = \pi \Ra^2 \rhoinf \vinf,
\eeq
where $\rhoinf$ is the upstream density  \citep{Hoyle:1939fl}. 
Flow convergence leads to shocks that imply a rate of dissipation of kinetic energy, or {\it drag luminosity},
\beq\label{ehl}
\ehl = \pi \Ra^2 \rhoinf \vinf^3  = \mhl \vinf^2,
\eeq
or a drag force of $F_{\rm d,HL} = \ehl /\vinf$ \citep[e.g.,][]{Iben:1993ka}. 

To estimate the  growth of the NS during the inspiral, we first approximate the inspiral timescale as 
\beq
t_{\rm insp} \approx  {E_{\rm orb} \over \ehl },
\eeq
where $E_{\rm orb }=  G M_{\rm NS} m / 2a $ and $m$ is the enclosed companion mass at a given orbital separation, $a$. The accreted mass is thus  $ \mhl t_{\rm insp}$, or
\beq
\Delta M_{\rm NS} \sim \mhl { E_{\rm orb} \over \ehl } \sim {\frac{ M_{\rm NS} m }{ 2 (M_{\rm NS} + m ) } }  
\eeq
where we further assume that $\vinf^2 = G (M_{\rm NS} + m) /a$ in the second equality.  This estimate reproduces, at the simplest level, the arguments of \citet{Chevalier:1993by} and later \citet{Brown:1995jj} and \citet{Bethe:1998jv} who argue that the NS should grow substantially during its inspiral. 

\subsection{Microphysics and Hypercritical Accretion}\label{sec:hyperaccretion}

The microphysics of accreting gas imposes several further scales on the accretion rate. 
The first of these is the Eddington limit. 
When the accretion luminosity reaches 
$L_{\rm Edd} = 4\pi G M_{\rm NS} c / \kappa$, where $\kappa$ is the opacity, radiation pressure counteracts gravity and halts the accretion flow. This limit on the accretion luminosity implies a limit on the accretion rate, 
\beq
\dot M_{\rm Edd} \approx 2 \times 10^{-8} \left(\frac{R_{\rm NS} }{12\text{km}} \right) \left(\frac{\kappa }{0.34\text{cm}^2 \text{ g}^{-1}} \right)^{-1} M_\odot \text{ yr}^{-1}.
\eeq

The Eddington limit may be exceeded under certain conditions if photons are trapped in the accreting flow and carried inward. Photons are trapped within the flow within a trapping radius of approximately \citep{Houck:1991kc},
\beq
R_{\rm tr} = \frac{\dot M \kappa}{4 \pi c} \approx  5.8 \times 10^{13} \left(\frac{\dot M }{M_\odot \text{yr}^{-1}} \right) \left(\frac{\kappa }{0.34\text{cm}^2 \text{ g}^{-1}} \right) \text{cm} .
\eeq
Because the NS has a surface, at small radii an accretion shock forms to stall the infalling gas. Within this shock neutrinos are the dominant cooling channel. The shock radius is
\beq
R_{\rm sh} \approx 1.6 \times 10^8 \left(\frac{\dot M }{M_\odot \text{yr}^{-1}} \right)^{-0.37} \text{cm},
\eeq
where the scaling with accretion rate arises from the  neutrino cooling function \citep{Houck:1991kc}. When $R_{\rm tr} > R_{\rm sh}$, accretion energy is advected into the neutrino-cooling layer and super-Eddington, or hypercritical, accretion can proceed \citep{Houck:1991kc}. This implies a lower-limit accretion rate of 
\beq
\dot M_{\rm hyper} \approx 1.9 \times 10^{-4} \left(\frac{\kappa }{0.34\text{cm}^2 \text{ g}^{-1}} \right)^{-0.73} M_\odot \text{yr}^{-1},
\eeq
where if $\dot M\gtrsim \dot M_{\rm hyper} \sim 10^4 \dot M_{\rm Edd}$ accretion can proceed despite the violation of the photon Eddington limit. No cooling solutions exist for $ \dot M_{\rm Edd} < \dot M < \dot M_{\rm hyper}$, so  mass  supplied at these rates  can only be incorporated at $ \dot M_{\rm Edd} $. 

Further investigation of these basic claims came in the form of multidimensional simulations, that confirmed that hypercritical accretion can reach a steady-state for some range of accretion rates while at others high entropy plumes intermittently overturn the flow \citep{Fryer:1996kr}. For a flow with some rotational support, the critical $\dot M_{\rm hyper}$ may be somewhat higher than for the spherical case described above \citep{Chevalier:1996kr,Brown:2000hm}. 

\subsection{Limits on the Accretion Rate due to Flow Asymmetry }\label{BHL}

\begin{figure*}[tbp]
\begin{center}
\includegraphics[width=0.40\textwidth]{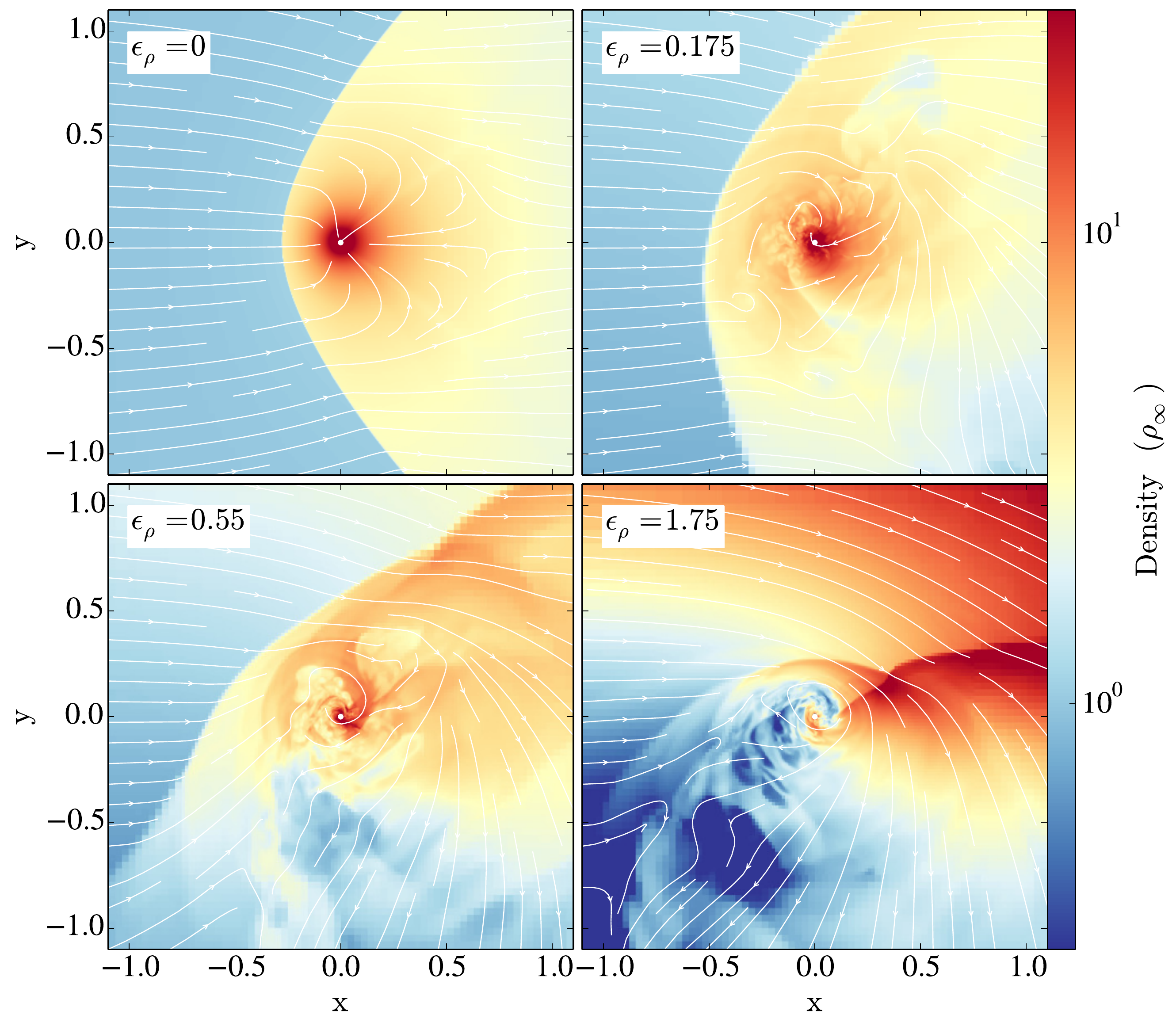}
\hspace{1.5cm}
\includegraphics[width=0.40\textwidth]{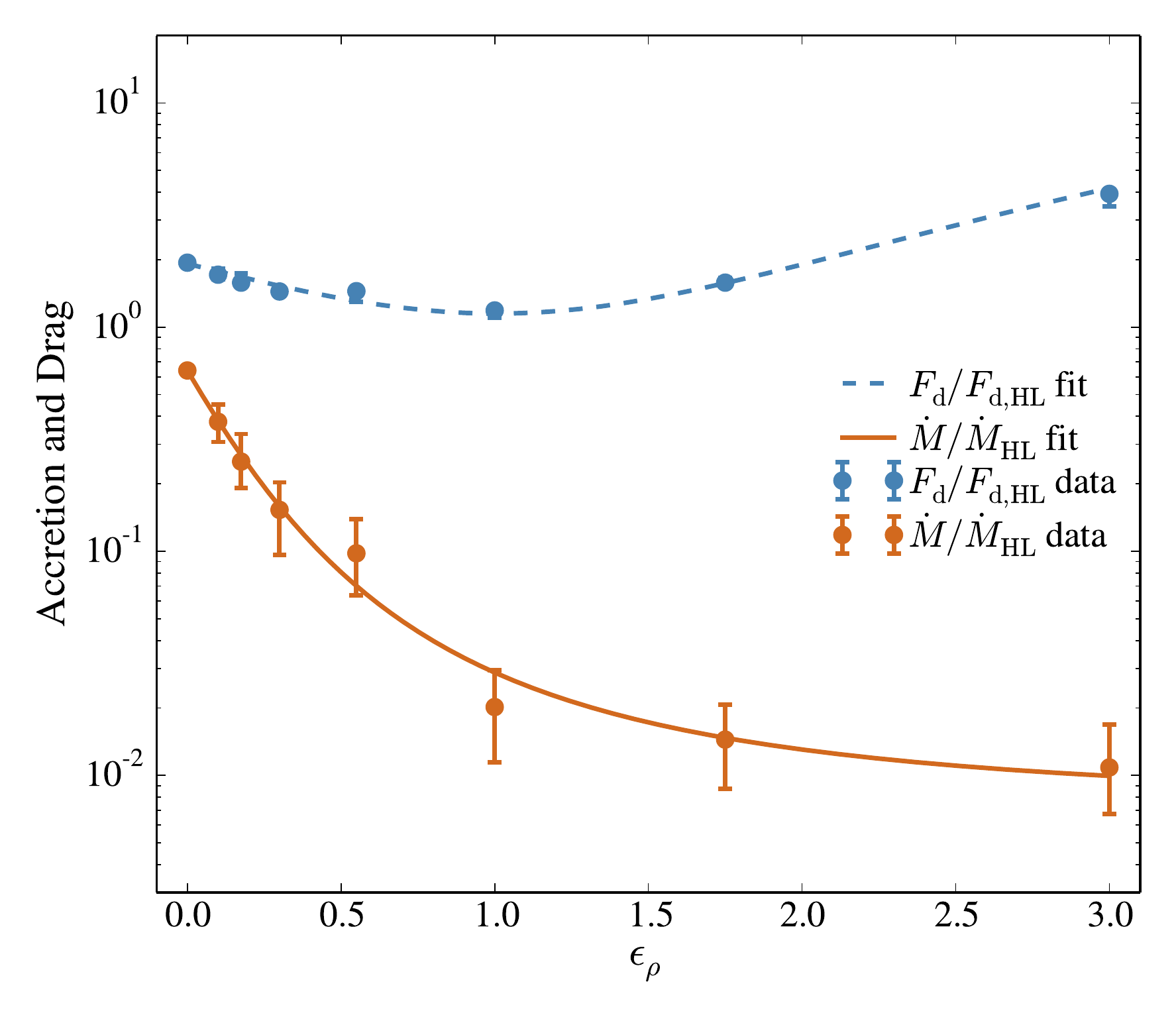}
\caption{Flow morphologies, drag, and accretion in 3D simulations of HLA with an upstream density gradient. The introduction of upstream density gradients, as found in CE evolution, breaks the symmetry of HLA and gives rise to the tilted bow shock structures seen here. The coordinates in the flow panel are in units of the accretion radius, $\Ra$, and the accretor is defined as an absorbing sink condition with $\Rs =0.01\Ra$ surrounding a central point mass. Flow momenta do not cancel in the wake of the accretor with upstream inhomogeniety and the post-shock region is defined by  rotation. In the right panel, we compare the resultant drag and accretion rate normalized to values anticipated by HLA.  We find that the drag force depends only mildly on density gradient, but the accretion rate decreases drastically as the density gradient, $\erho$, steepens. These calculations use a gamma-law equation of state with $\gamma=5/3$. }
\label{fig:bhl}
\end{center}
\end{figure*}

In order to track accretion onto a NS during CE, we need a clear prediction of the accretion rate as a function of CE structure. 
In \citet{MacLeod:2014tn}, we use the FLASH adaptive mesh hydrodynamics code \citep{Fryxell:2000em} to extend three-dimensional (3D) simulations of HLA to consider the role of an inhomogeneous upstream medium. We characterize the density gradient across the accretion radius as
\beq
\erho = {\Ra \over H_{\rho}},
\eeq
where $H_\rho = - \rho dr/d\rho$, the density scale height. A planar density gradient is then applied along the simulation $y$-axis, perpendicular to the direction of motion, with $\rho = \rho_\infty \exp ( \erho y)$, where $\rho_\infty$ is the density at zero impact parameter, $y=0$.  We find that typical values for $\erho$ in CE range from $\erho \approx 0.3-3$. 

Strong density gradients break the symmetry that defines HLA, severely limiting accretion. The momenta of opposing streamlines do not fully cancel with the introduction of inhomogeneity, and the resulting flow carries angular momentum with respect to the accretor. Thus, even if material is gravitationally captured it may not be accreted because of this angular momentum barrier. The HLA formula, Equation \eqref{mhl},  drastically overestimates the resultant accretion rate \citep[see also][]{Ricker:2007kc,Ricker:2012gu}. In Figure \ref{fig:bhl}, we show how  flow morphology, drag force, and accretion rate change with steepening density gradients. 

\section{Inspiral and Accretion}\label{sec:inspandacc}
In order to trace the NS inspiral through the dynamical phase of CE evolution, we integrate coupled equations for the evolution of the orbit and accretion onto the NS. We discuss our initial models, evolution equations, and findings  below. 

\subsection{Methods}\label{sec:methods}

\begin{figure}[tbp]
\begin{center}
\includegraphics[width=0.49\textwidth]{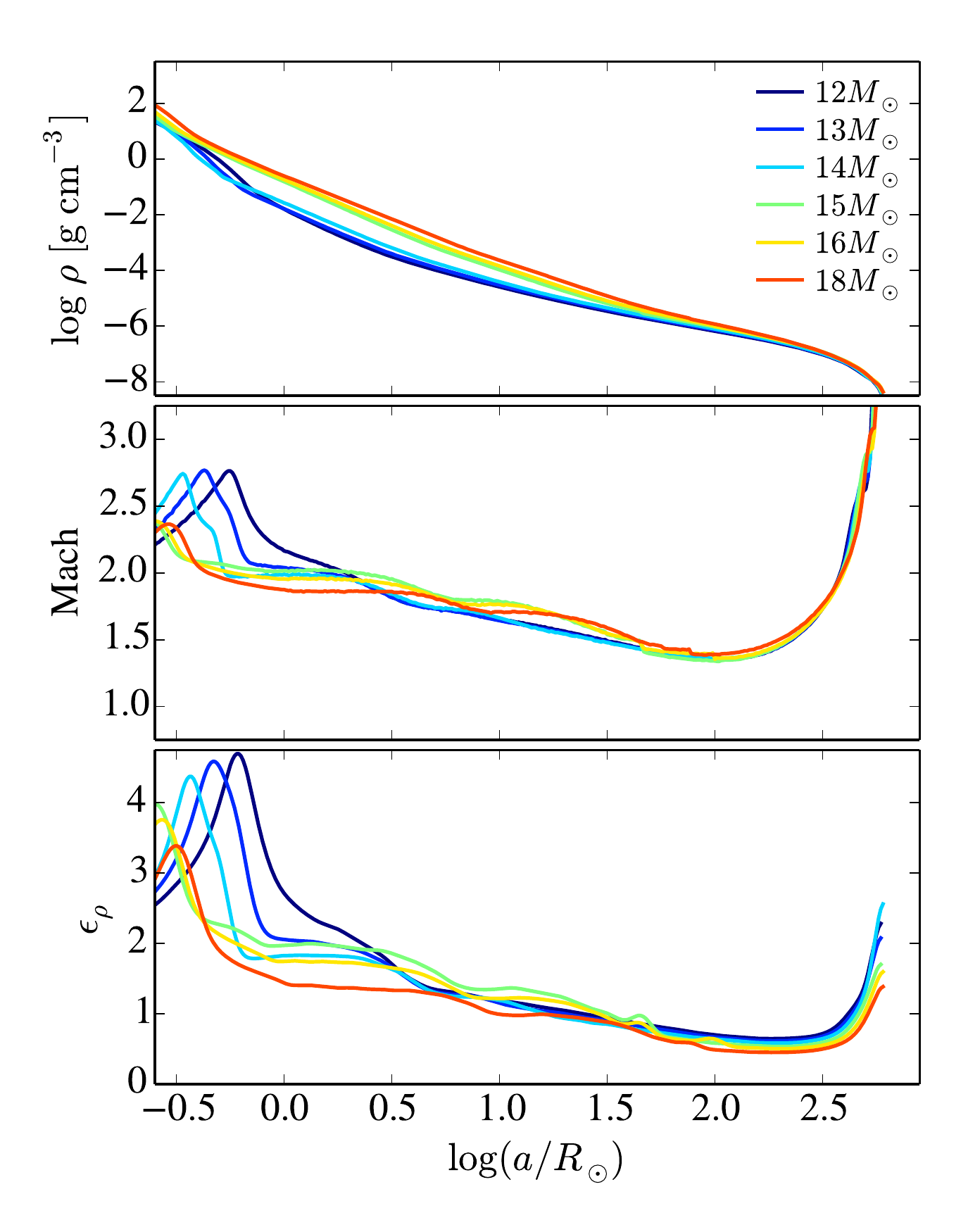}
\caption{Envelope structure of a range of giant star models that share $600R_\odot$. The top panel shows the density profile of the envelopes, while the center and lower panels show flow the Mach number and the density gradient that would be experienced by an inspiralling $1.33 M_\odot$ NS. For much of the stellar interior, mach numbers are moderate $\mach \approx 1.5-3$ with density gradients of $\erho \approx 1-2.5$, representing substantial density inhomogeneity across the accretion radius. Spikes in the density gradient are seen in the deep interior at transitions in chemical composition.  }
\label{fig:structure}
\end{center}
\end{figure}

To create approximate CE conditions, we evolve single stars in the MESA stellar evolution code \citep[version 5527:][]{Paxton:2011jf,Paxton:2013th}. During the giant-branch phase, a CE event may be initiated when the radius of the stellar envelope grows to be similar to the binary separation, $R_* \sim a$. 
We make the simplifying approximation of a static CE profile. This is most valid when the companion mass is much greater than the NS mass, $M_{\rm comp}\gg M_{\rm NS}$. 
The progenitors of NSs in binaries are massive stars, so we calculate the structure of giant-branch envelopes of with initial masses of $12-20 M_\odot$. 
A comparison of these envelope structures, and the typical flow Mach numbers and density gradients they give rise to, is shown in Figure \ref{fig:structure}.

Orbital energy is dissipated at a rate 
\beq
\dot E_{\rm orb} = - F_{\rm d}(\erho) \vinf,
\eeq
where $F_{\rm d}(\erho)$ is approximated using a fit to our simulation results described in Section \ref{BHL},
\beq\label{dragnumerical}
\frac{F_{\rm d}(\erho) }{F_{\rm d,HL}} \approx f_1 + f_2 \erho + f_3 \erho^2 ,
\eeq
 with $f_i = (1.91791946, \ -1.52814698, \  0.75992092)$.
As a result of this drag, the orbital separation evolves at a rate
$
\dot a = \dot E_{\rm orb} (da/dE_{\rm orb})
$.
We terminate our integration of the dynamical inspiral when the integrated change in orbital energy equals the envelope binding energy at a given CE  radius $\Delta E_{\rm orb}(a) = E_{\rm env}(a)$, equivalent to $\alpha_{\rm CE}=1$, \citep{Webbink:1984jd}.  The envelope binding energy is computed as 
\beq
E_{\rm env}(a) = \int_{m(a)}^M u-\frac{G m}{r} dm,
\eeq
where we have included both the gravitational binding energy and internal energy of the stellar fluid.

The expression regulating accretion onto the NS is 
\beq
\dot M_{\rm NS}  = \dot M(\erho),
\eeq
where, like the drag, $\dot M(\erho)$ is a fit to our numerical results, 
\beq\label{mdotnumerical}
\log \left(\dot M(\erho)/\mhl \right) \approx m_1 + \frac{m_2}{1 +m_3 \erho + m_4 \erho^2},
\eeq
with $m_i = (-2.14034214, \ 1.94694764, \ 1.19007536,\ 1.05762477)$. To give a baseline for comparison, we also compute orbital inspiral sequences using HLA theory, with $\dot E_{\rm orb} = \ehl$ and $\dot M_{\rm NS} = \mhl$. 

We make several approximations that likely lead our calculation of the accreted mass in CE to be an overestimate. First, in assuming a static structure for the CE, we may overestimate the local density of the dispersing envelope \citep[see, e.g.,][]{Ricker:2012gu}. Second, we assume that hypercritical accretion and cooling by neutrinos are effective above $\dot M_{\rm hyper}$, despite the fact that this may not apply at all values of $\dot M > \dot M_{\rm hyper}$, in particular with varying amounts of angular momentum \citep{Fryer:1996kr,Chevalier:1996kr,Brown:2000hm}.
Finally, we compute the mass accretion rate, Equation \eqref{mdotnumerical}, assuming that all mass passing through $\Rs=0.01\Ra$  is able to propagate the additional 2-3 orders of magnitude in radial scale to $R_{\rm sh}$, where cooling can occur. 
Thus, our integration represents an upper limit for the potential accreted mass onto an embedded NS.

\subsection{Results}\label{sec:results}

\begin{figure}[tbp]
\begin{center}
\includegraphics[width=0.45\textwidth]{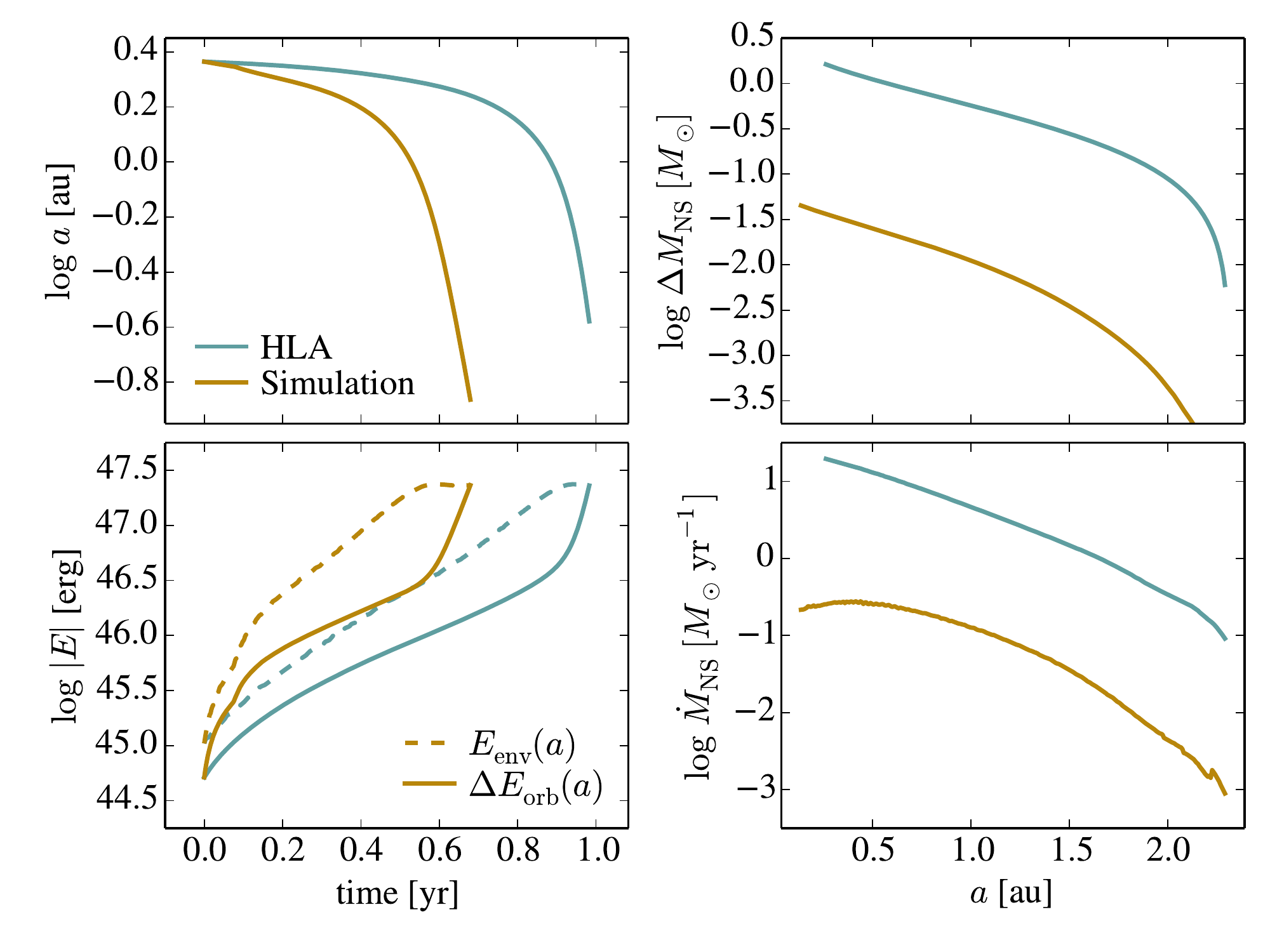}
\caption{Orbital inspiral of an originally $1.33M_\odot$ NS through the envelope of its $12 M_\odot$, $500R_\odot$ companion. The left panels show evolution of the orbital separation (top), and energies (bottom). The right hand panels show the mass accretion of the NS in terms of the separation. Blue lines show the results assuming HLA, while yellow lines take into account the effect of asymmetry on the accretion rate. In HLA, the NS grows well beyond the maximum NS mass, acquiring more than a solar mass during its inspiral. However, the loss of symmetry in the accretion flow limits $\dot M_{\rm NS}$, such that the NS gains less than $0.1 M_\odot$ and survives the CE.    }
\label{fig:singleinspiral}
\end{center}
\end{figure}

We begin by comparing orbital inspirals based on HLA theory with simulation coefficients for drag and accretion in Figure \ref{fig:singleinspiral}. This comparison highlights the need to consider the role of the structure of the CE around the embedded NS.  
In the HLA case, the NS gains more than $1M_\odot$, enough mass to push it above the $\sim 2M_\odot$ maximum NS mass, and in agreement with our analytic prediction of Section \ref{sec:CE_HLA}. 
However, in the simulation case, we see that $\dot M_{\rm NS}$, and in turn $\Delta M_{\rm NS}$ are both severely limited by flow asymmetry. 
$\dot M_{\rm NS}$ is still sufficiently high to allow hypercritical accretion to proceed \citep{Chevalier:1993by}, but the NS gains less than $0.1M_\odot$ during its inspiral. This accreted mass represents a few percent of the NS's mass. Thus, the final compact object is a slightly more massive NS, rather than a BH.

\begin{figure}[tbp]
\begin{center}
\includegraphics[width=0.45\textwidth]{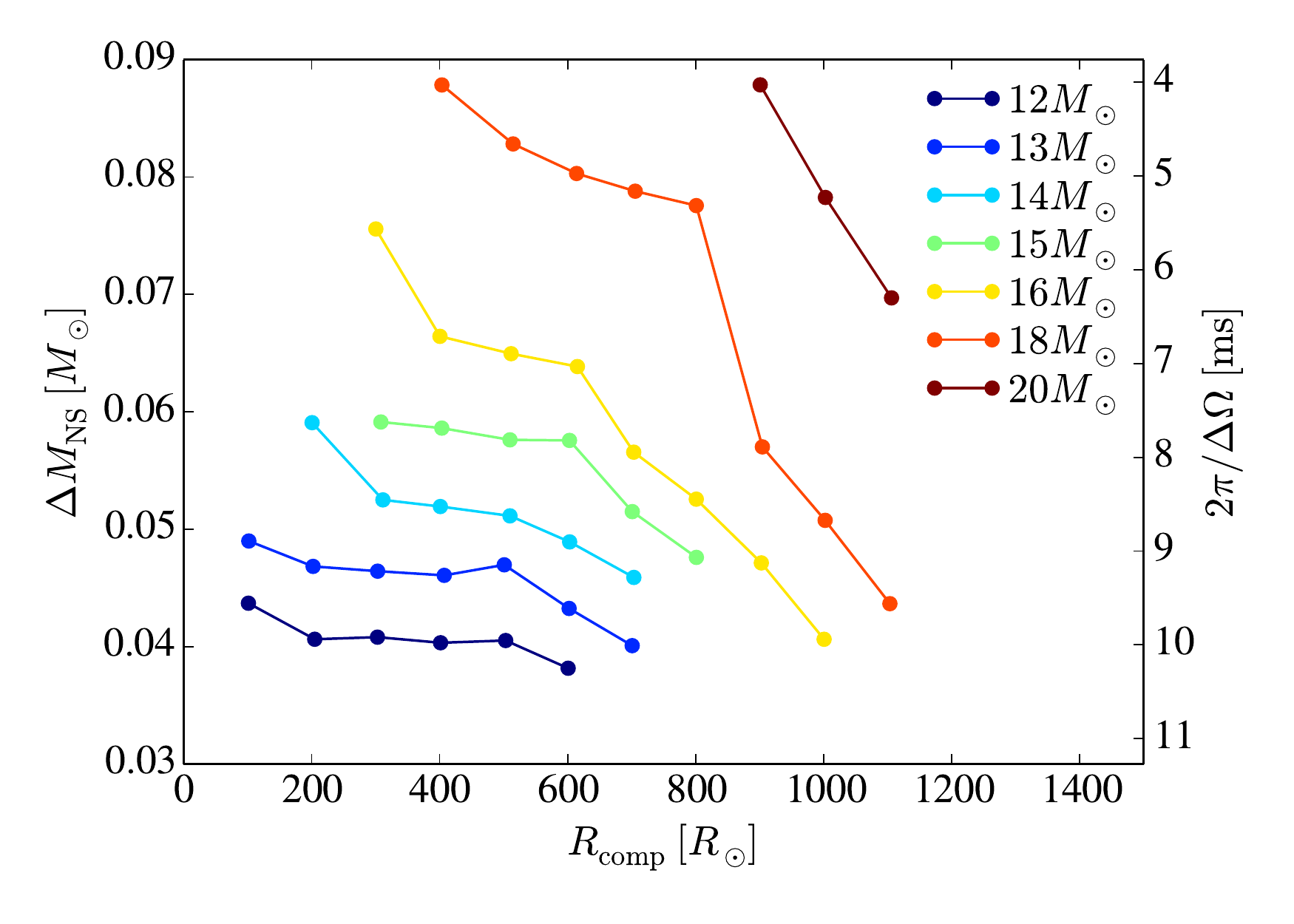}
\caption{Post-CE states for originally $1.33M_\odot$ NSs involved in interactions with wide variety of companions. Companions range in pre-CE mass from $12-20M_\odot$, in radius from $100-1100 R_\odot$,  and have convective envelopes. CE events initiated with smaller-radius companions than those plotted (for a given mass) result in merger rather than envelope ejection. All evolutions computed result in NSs surviving CE, with none expected to undergo accretion-induced collapse to a BH. NSs generally gain more mass in interactions with more massive companions, but very extended radius at the onset of CE can lead to less mass accumulation. To compute the right-hand axis we assume material is accreted from a Keplerian disk, and that adopt median NS properties of  $M_{\rm NS}\approx1.39M_\odot$, $R_{\rm NS}=12$km.  NSs undergoing these CE episodes survive to find themselves in close partnerships with helium-rich companions and orbital periods ranging from hours to days. 
}
\label{fig:all}
\end{center}
\end{figure}

We now extend our calculation to consider a diversity of pre-CE structures. 
 In Figure \ref{fig:all}, we plot only those structures for which the CE ejection is  successful,  where $\Delta E_{\rm orb}(a) = E_{\rm env}(a)$  at some radius in the stellar interior. In general, this criteria is satisfied when a distinct helium core and convective envelope structure forms. Minimum orbital periods are in the range 0.1-2 yr as determined by the masses and radii of the companions at the onset of CE. This is in agreement with the analysis of \citet{1995ApJ...445..367T} who found a dividing period of 0.8-2 yr for companions of 12-24 $M_\odot$. CE events involving less-evolved  giants (smaller $R_{\rm comp}$ than those plotted) likely lead to complete mergers because the NS is unable to eject its companion's envelope. The merger could result in the formation of either stably burning  Thorne-Zytkow objects \citep{1977ApJ...212..832T,2014MNRAS.443L..94L} or explosive transients \citep{2013ApJ...764..181F}.

In each CE structure considered,  the NS survives CE. Perhaps more strikingly, it gains only a few percent of its own mass across a broad array of different envelope structures. 
In general, NSs gain the most mass in interactions with more massive companions. There are two reasons for this effect. First, the NS must spiral deeper to eject its companion's envelope when the mass ratio is larger \citep{Webbink:1984jd}. Second, large mass ratios imply that $\Ra$ is a smaller fraction of $R_{\rm comp}$, and as a result, the effective density gradient, $\erho$, is reduced (Figure \ref{fig:structure}), allowing for more efficient accretion (Figure \ref{fig:bhl}). The NS gains less mass in interactions with more extended companions (for a given mass) because these envelopes are comparatively easier to unbind.

Mass accretion implies a spin-up of the NS based on the specific angular momentum of accreting material. In cases where the NS mass is not well determined, the pulsar spin period can still offer constraints on the accreted mass. 
The spin-up is $ \Delta \Omega  = \Delta L / I_{\rm NS}$, where $\Delta L$ is the accreted angular momentum and $I_{\rm NS}$ is the NS's moment of inertia. 
We estimate $\Delta L$ assuming that material is accreted from a neutrino cooled disk surrounding the NS, $\Delta L \approx \Delta M \sqrt{G M_{\rm NS} R_{\rm NS}}$, and that  $I_{\rm NS} =2/5 M_{\rm NS} R_{\rm NS}^2$. The post-CE spin period of the recycled NS can then be estimated as $2\pi / \Delta \Omega$. 
The calculated spin periods are shorter than those observed presently for first-born pulsars in double NS systems \citep[$P \sim 20-100$ ms,][]{2011MNRAS.413..461O}. However, NSs with magnetic fields $\gtrsim 10^{10}$G will  quickly spin down from these initial periods, so spin constraints are most valuable where the NS magnetic field is small (and thus the spin-down timescale is long). Two of the ten pulsars listed by \citet{2011MNRAS.413..461O} meet this criterion, pulsars J1518+4904 and J1829+2456. Neither of these object has a well-determined mass, but their measured pulse periods (both $\approx41$ms) and period derivatives  $\lesssim 10^{-19} s s^{-1}$ imply spin down timescales $>10$ Gyr, allowing direct comparison to Figure \ref{fig:all}. 
If spun-up by accretion during CE, the spin of these objects is consistent with having gained of order $0.01 M_\odot$.

\section{Discussion}\label{sec:discussion}
We have self-consistently evaluated the mass growth of a NS embedded within a CE, 
taking into account that the accretion rate depends sensitively on the structure of the CE \citep{MacLeod:2014tn}. 
Although our integration likely represents an upper limit (as discussed in Section \ref{sec:methods}), we observe that NSs gain only a few percent of their own mass during CE episodes. These objects emerge from CE  only mildly heavier and more rapidly spinning, rather than undergoing accretion-induced collapse to BHs \citep{Chevalier:1993by,Armitage:2000eg}. It appears that density gradients in typical CE structures are the missing link needed to reconcile theories of hypercritical accretion onto NSs \citep{Houck:1991kc,Chevalier:1993by,Fryer:1996kr} with the narrow observed mass distribution of NS masses \citep{Schwab:2010du,Ozel:2012cv,Kiziltan:2013ky}. 
This result hints that forming double NS binaries may not require a finely-tuned evolutionary channel \citep{Brown:1995jj,Bethe:1998jv}, but they could instead emerge from within the standard CE binary evolution framework \citep[e.g.,][]{Stairs:2004go,2006csxs.book..623T}. 

Further investigation is certainly needed to probe the efficiency of CE ejection by embedded NSs, as well as the dynamical timescale effects of envelope dispersal \citep[for example, as studied by ][]{1995ApJ...445..367T}. 
We note that a reduction in $\dot M$ , as compared to $\mhl$, may hinder envelope ejection in that any potential accretion disk feedback \citep{Armitage:2000eg,Papish:2013tf} would be weakened relative to the envelope's binding energy. This hints that other forms of feedback that may be less dependent on $\dot M$, like nuclear burning or recombination, may be of more assistance in CE ejection \citep{Iben:1993ka,Ivanova:2014wq}.  
Studies that consider these energy sources can best determine the critical separation (or orbital period) that divides binaries that merge from those that successfully eject their envelopes. 

The  CE stage described here is not the full story of the evolution of a binary. 
 In many binaries, the first-born pulsars  interact with their helium-star companions following the CE  \citep{2006csxs.book..623T}. If an additional CE were to occur, the low mass and steep density gradients of the He star's typically radiative envelope \citep{2002MNRAS.331.1027D} suggest a relative ease of envelope ejection and a low  accretion efficiency. 
However, as the most-recent interaction, this phase of mass transfer or CE  would  be responsible for the current spin of the first-born pulsar.  
These more complex interaction histories are best traced with population synthesis calculations, where the ramifications of observed masses, spins, and orbital eccentricities offer a window to the outcome of the CE phase \citep{Kalogera:2007kh,Dominik:2012cw}.

We anticipate that moving beyond the energy formalism of CE \citep{Webbink:1984jd} to also consider CE structure, as parameterized by density gradient $\erho$, will shape the channels through which double compact binaries can be expected to form.  
 As a result of the structures of their companions,  few to none of the NSs entering CE episodes should be expected to collapse to BHs.

\begin{acknowledgements}
We thank the anonymous referee for thoughtful and constructive feedback and C. Fryer, J. Guillochon, V. Kalogera, B. Kiziltan, D. Lin, A. Loeb, S. de Mink, R. Narayan, F. \"Ozel, P. Podsiadlowski, M. Rees, T. Tauris, and S. Woosley for guidance and helpful discussions. We  acknowledge support from the David and Lucile Packard Foundation, Radcliffe Institute for Advanced Study, NSF grant AST-0847563, and the Chancellor's Dissertation-Year Fellowship at UCSC. 
\end{acknowledgements}

\bibliographystyle{apj}

\end{document}